# Inversion Framework of Internal Mass Distribution Parameters of Asteroid Apophis from Dynamical Observations

**Yiting Li[a], Chenyang Huang[b], Yang Yu[c*], Shengping Gong[d], He Zhang[e]**

[a]School of Shen yuan, Beihang University, No.37 Xueyuan Road, Haidian District, Beijing 100191, China. Email:2176374479@qq.com
[b]Department of Aerospace Science and Technology, Politecnico di Milano, Via La Masa 34, 20156 Milano, Italy. Email:chenyangh1314@gmail.com
[c]School of Aeronautic Science and Engineering, Beihang University, No.37 Xueyuan Road, Haidian District, Beijing 100191, China. Email:yu.yang@buaa.edu.cn
[d]School of Astronautics, Beihang University, No.37 Xueyuan Road, Haidian District, Beijing 100191, China. Email:gongsp@buaa.edu.cn
[e]Beijing Institute of Spacecraft System Engineering, No.104, Youyi Road, Haidian District, Beijing 100094, China. Email:Zhanghe_maoqiu@163.com

**Abstract**

The detection of the internal mass distribution of asteroids is of great significance for understanding their origin, evolution, and mission planning for exploration. Previous approaches rely on indirect density estimates or close-range spacecraft gravity inversion, which have limited applicability. This paper presents a proof-of-concept framework to infer the internal mass properties of asteroid (99942) Apophis during its close Earth flyby in 2029 using dynamical observations collected during the encounter. We establish a dynamical mapping from the evolution of orbital and rotational states to internal structural parameters, formulate it as an inverse problem, and solve it using Particle Swarm Optimization. The algorithm is first validated on a regular ellipsoidal model and then applied to three mass distribution models based on the actual shape of Apophis. Under ideal observation conditions, the relative error of the inverted moment of inertia ratios can be below 0.001%, and the absolute error of the center-of-mass position reaches the order of $10^{-5}$ meters. The algorithm successfully distinguishes among different internal structures. When realistic measurement noise is introduced, the inversion accuracy degrades. A sensitivity analysis reveals that the accuracy of the inertia tensor inversion is primarily limited by angular velocity measurement noise, whereas center-of-mass determination is highly sensitive to the precision of position and velocity data. This study provides a proof-of-concept for a technically feasible and cost-effective approach to infer asteroid internal structure during close encounters, and also highlights the critical data accuracy requirements for practical application, offering guidance for future observation campaigns.

## 1 Introduction

With the rapid advancement of deep-space exploration missions, small solar system bodies have become a focal point of research in aerospace technology and planetary science over recent decades, their scientific value and strategic significance reaching unprecedented levels. Asteroids, as primordial remnants from the era of planet formation, hold vital clues to the early history of the solar system (Izidoro et al. 2016; Badescu V et al. 2013). Their internal structures, particularly their mass distribution, are not only critical for planning near-Earth exploration missions but also offer key insights into their formation, current state, and evolution. However, owing to their vast distances and weak gravitational fields, directly probing the internal structures of these bodies remains a major challenge in planetary science.

Given this, researchers have long depended on indirect methods to infer asteroid internal structures. Traditional approaches typically derive three-dimensional shape models from radar or optical observations, estimate the total mass via orbital perturbations or binary system dynamics, and then compute a bulk density (Carry et al. 2012). While straightforward, this method essentially assumes a uniform internal density distribution, making it difficult to reveal potential complexities, as asteroid interiors may exhibit significant mass heterogeneity.

To overcome this limitation, more sophisticated indirect methods have been developed, particularly in the context of dedicated spacecraft missions. Scheeres et al. (2020) estimated the density distribution of asteroid Bennu by utilizing particle ejection events observed on its surface by the OSIRIS-REx spacecraft, combined with direct tracking via deep-space network and spacecraft-based optical imaging, as well as morphology-driven methods and global gravity inversion techniques. Konopliv et al. (2011) and Ermakov et al. (2014) processed X-band Doppler tracking and optical image tracking data from the DAWN mission to derive spherical harmonic models of the gravitational fields of Ceres and Vesta, and subsequently inferred their internal structural parameters by incorporating prior information from planetary geology. Although such close-range spacecraft measurements can invert the gravity field with relatively high accuracy and thereby constrain internal structure, they incur substantial mission costs and rely on complex data processing, and are applicable only to a limited number of pre-selected targets. Therefore, a method capable of relatively rapid, accurate, and low-cost inference of internal structure from routine observational data is highly desirable.

A unique natural experiment that could help fill this gap is the close flyby of an asteroid past a massive planet. During such an event, the differential gravitational force (tidal force) of the planet exerts a torque on the asteroid, perturbing both its orbit and, more importantly, its rotational state. Crucially, this dynamical response is governed by the asteroid's low-order mass distribution parameters: its center of mass and its inertia tensor. Therefore,by analyzing these observed changes, one can, in principle, construct an inverse problem to infer the very internal parameters that control the response. Asteroid (99942) Apophis provides an exceptional opportunity to test this concept. It is predicted to make a close flyby of Earth in April 2029. According to current JPL ephemeris predictions, it will approach to within approximately 31,600 kilometers from the Earth's surface, which is lower than the orbital altitude of some geosynchronous satellites (JPL Small-Body Database Browser, 2022). This exceptionally close encounter will subject the asteroid to significant gravitational perturbations and tidal torques from Earth, leading to noticeable alterations in its orbital trajectory and rotational state (Benson et al., 2023). Previous studies have primarily focused on estimating the total mass of a target asteroid through gravitational interactions with other celestial bodies (Scheeres et al., 2015; Zhang et al., 2021).This prompts the question of whether we can go a step further: can we directly invert its internal mass distribution by analyzing the observed changes in its motion state parameters—such as position, velocity, attitude, and spin—during the flyby? However, determining the internal mass distribution uniquely remains highly challenging due to the strongly nonlinear dynamical environment of the asteroid and the inherent degeneracy in internal structure models. Different density distributions can lead to the same dynamical state parameters (Otegi et al. 2020).

From a dynamical perspective, the internal mass distribution of a celestial body can be fully characterized by its first and second-order moments (center of mass and moment of inertia tensor, respectively). This paper proposes an identification method for the inertia products and the center-of-mass position of an asteroid based on rigid-body dynamics equations and a least-squares optimization framework. The aim is to fully utilize the observational opportunity presented by the 2029 Apophis flyby to infer its key internal mass parameters. We establish a mapping relationship between the evolution of the asteroid's orbital and spin states during the flyby and its internal parameters. The inertia matrix, including off-diagonal terms, and the center-of-mass position within the asteroid's body-fixed frame are treated directly as the parameters to be solved. A heuristic algorithm (Particle Swarm Optimization) is employed to address this inverse problem. Test results demonstrate that this method effectively ensures global convergence and mitigates the impact of local minima.To validate the proposed method, we constructed a digitally-twinned dynamical model based on the currently predicted Apophis flyby scenario and anticipated observational conditions. An uncertainty and sensitivity analysis of the observations was also conducted.

The structure of this paper is organized as follows: Section 2 details the 2029 flyby scenario of asteroid Apophis past Earth and the expected observable data, followed by an explanation of the inversion method’s fundamental principles and the particle swarm optimization algorithm. Section 3 presents the results, first validating the method’s effectiveness through an analytical model, then demonstrating its application to asteroid Apophis, including a rigorous uncertainty analysis. Section 4 discusses the significance of this study, the scientific and engineering value of the method, and its application prospects for the upcoming historic flyby event.

## 2 Methods

### 2.1 The 2029 Apophis Encounter Scenario

We adopt the predicted close flyby of near-Earth asteroid (99942) Apophis on April 13, 2029, as our baseline scenario. The extremely small geocentric distance during this flyby is expected to cause significant perturbations to the asteroid's orbital and spin states due to Earth's gravitational gradient. The aim of this work is to estimate the asteroid’s internal structural parameters using the changes in its orbital and rotational state expected to be observed during the flyby. According to JPL ephemeris predictions, Apophis will pass Earth on April 13, 2029, at 21:46 UT at a distance of approximately 31,600 km from the Earth's surface. Its velocity relative to Earth upon approach is 6.0 km/s. At the point of closest approach, Earth's gravity will accelerate it to 7.4 km/s, before it decelerates back to 6 km/s as it departs (Pravec et al. 2014). The specific changes in its orbital parameters before and after the flyby are shown in Table 1.

| | Orbit Type | Orbital Period (days) | Semi-major Axis (AU) | Perihelion (AU) | Aphelion (AU) | Inclination (°) | Eccentricity |
|---|---|---|---|---|---|---|---|
| **Before Flyby** | Aten | 323.6 | 0.922 | 0.746 | 1.10 | 3.34 | 0.191 |
| **After Flyby** | Apollo | 423.1 | 1.103 | 0.895 | 1.31 | 2.22 | 0.189 |

**Table 1.** Orbital parameters of Apophis before and after the flyby. Data sourced from the JPL Horizons ephemeris system.

Existing observations suggest that Apophis likely has an elongated bilobate structure, with dimensions of approximately 450×170 meters (Brozović et al., 2017). Since Earth’s tidal forces dominate during the brief flyby, we simplify the Earth as a point-mass gravitational source and neglect perturbations from other bodies such as the Sun and the Moon.

**2.2 Expected Observational Constraints**

In practical scenarios, observational parameters are often non-ideal, which inevitably introduces disturbances into the inversion results of the output parameters. To evaluate the actual performance of our proposed method, we introduce Gaussian noise into the input data of the inversion model. The setting of observational accuracy references the technical precision of past observations of near-Earth objects, primarily based on the capabilities of radar and optical facilities.

Radar observation is currently the most widely used technique in planetary science for this purpose. By actively emitting electromagnetic waves to detect a target, they can precisely determine its orbit, rotation, and topography(Virkki et al., 2023). Currently, radar ranging and Doppler velocimetry for near-Earth asteroids can achieve relative accuracies better than $1\times10^{-7}$, providing crucial support for planetary science and planetary defense (Naidu et al., 2016). Ground-based radar systems commonly used internationally for observing near-Earth objects mainly include the Arecibo radio telescope in Puerto Rico and the Goldstone Deep Space Communications Complex in California. Since the 1970s, these systems have observed hundreds of near-Earth asteroids. The 34-meter DSS-13 antenna at Goldstone is equipped with an 80-kilowatt transmitter operating at a frequency of 7190 MHz. The high signal-to-noise ratio of this transmitter enables delay-Doppler imaging with a ranging resolution as fine as 1.875 meters, which is twice the finest ranging resolution achievable by the DSS-14 antenna (Naidu et al., 2016).

Space-based optical observation systems typically employ passive detection methods. For instance, the JWST-MIRI mid-infrared instrument can determine the right ascension and declination of celestial bodies with an absolute positioning accuracy better than 0.05 arcseconds (Ioannis et al., 2023). By processing consecutive multi-frame image positions, it can measure motion velocities with an accuracy on the order of arcseconds per hour. Compared to visible light, infrared observations offer stronger detection capabilities for faint celestial objects and can obtain richer spectral information about asteroids. Therefore, infrared-band telescopes are the preferred choice for space-based observation systems. Launched by the United States in 2009, NEOWISE is currently one of the primary international space-based asteroid observation systems. For target objects with a magnitude of 10, its positioning accuracy can reach 50 milliarcseconds, and its proper motion measurement accuracy ranges from 10 to 30 milliarcseconds per year (Mainzer et al., 2014).

The rotational period and attitude orientation of an asteroid can be initially constrained using optical imaging or delay-Doppler data, followed by the determination of parameter ranges through iterative or inversion algorithms combined with high-precision radio tracking. Based on previous high signal-to-noise studies of near-Earth asteroids, we conservatively estimate that the relative measurement accuracy of the rotational rate $\boldsymbol{\omega}$ is on the order of $10^{-7}$. For example, during the NEAR Shoemaker mission in orbit around 433 Eros, the orientation of the rotational axis (right ascension $11.3692 \pm 0.003°$, declination $17.2273 \pm 0.006°$) was determined using the PCODP method by tracking the motion of the same surface features across multiple images as Eros rotated, integrating radio tracking, optical imaging, and laser altimetry data. Furthermore, by processing long-term radio tracking data (ODP) and combining it with the spacecraft's orbital position and camera pointing model, the rotational rate was determined as $\omega$=1639.38922 ± 0.00015 °/day (Miller et al., 2002). The observational accuracy of these parameters provides a reference for setting noise in simulated data and evaluating the performance of the inversion method.

## 2.3 Forward Dynamics Model Setup

Most asteroids, having formed by gravitational accumulation over long evolutionary timescales, are thought to possess rubble-pile structures with non-uniform mass distributions. In our research, we found that the commonly used polyhedron method might lead to divergence in dynamical integration due to numerical errors at the altitude of Earth's orbit. Therefore, this paper adopts the mass-point cloud method to calculate the gravitational force exerted by Earth on the asteroid. The mass-point cloud method is relatively simple in principle, easy to implement, and capable of flexibly describing complex shapes and internal mass distributions (Werner and Scheeres, 1996).

To establish a tractable yet representative forward model, this study relies on a set of key assumptions.Their justification stems from the short timescale of the flyby and the dominance of Earth's gravity; potential impacts on the inversion results are acknowledged and discussed as limitations and future work.

**1.Two-Body Problem:** The dynamical system is simplified to the asteroid and Earth as a point mass. Perturbations from the Sun, Moon, and other celestial bodies are neglected, as the Earth's influence is expected to be dominant during the short, ~1-day flyby window.

**2.Rigid Body:** Apophis is assumed to be a rigid body. Although it may possess a rubble-pile structure, internal energy dissipation or deformation is likely negligible over the flyby timescale.

**3.Keplerian Orbit for COM:** The translational motion of the asteroid's center of mass is assumed to follow a Keplerian orbit around Earth. This implies that the attitude motion and the center of mass offset do not couple back into the translational dynamics. This is a reasonable first-order approximation given that the asteroid's dimensions are much smaller than the flyby distance.

**4.Known Shape Model:** A high-resolution three-dimensional shape model of Apophis is assumed to be available from pre-flyby radar and optical observations (Brozović et al., 2018). This model provides the geometric framework for distributing mass points.

Under these assumptions, we first establish the following coordinate systems to describe the dynamics of an asteroid's flyby past Earth:

**1.Geocentric Inertial Frame OXYZ:** The origin is located at the Earth's center. At the initial epoch, the line connecting the geocenter and the asteroid's center of mass is defined as the OX axis. The OZ axis is perpendicular to the asteroid's initial orbital plane; and the OY axis completes the right-handed system.

**2.Translational Frame Oxyz:** The origin is at the asteroid's center of mass. Its coordinate axes align with those of the inertial frame at the initial moment.

**3.Body-Fixed Frame Ox'y'z':** The origin is at the asteroid's center of mass. The coordinate axes point along the asteroid's three principal axes of inertia, defined in order of increasing moment of inertia as the x-axis (minimum inertia), y-axis (intermediate inertia), and z-axis (maximum inertia).

During the approximately one-day flyby period, it is assumed that the asteroid is subject only to Earth's gravitational force, with other external perturbations neglected. We discretize the asteroid into N mass elements (mass points), assigning each point a different mass weight $m_i$ according to the asteroid's shape and structure. The sum of all point masses equals the asteroid's total mass M.Let $\boldsymbol{r_i}$ be the position vector of the i-th mass point in the inertial frame OXYZ relative to the Earth's center.The gravitational force $\boldsymbol{F_{ext}}$ acting on the asteroid in the inertial frame OXYZ and the gravitational torque $\boldsymbol{\tau_{ext}}$ about its center of mass in the body-fixed frame Ox'y'z' are obtained by summing the contributions from all mass points:

$$\boldsymbol{F_{ext}} = \sum_{i=1}^{N} \frac{GMm_i(\boldsymbol{r_c} + \boldsymbol{R_{BI}} \cdot \boldsymbol{\rho_i})}{\left|\boldsymbol{r_c} + \boldsymbol{R_{BI}} \cdot \boldsymbol{\rho_i}\right|^3} \tag{1}$$

$$\boldsymbol{\tau_{ext}} = \sum_{i}^{N} \boldsymbol{\rho_i} \times (\boldsymbol{R_{BI}^T} \cdot \boldsymbol{F_i}) \tag{2}$$

Here,$GM$ is Earth's gravitational parameter, $\boldsymbol{r_c}$ is the position vector of the asteroid's center of mass in the inertial frame, $\boldsymbol{\rho_i}$ is the position vector of the i-th mass point in the body-fixed frame, and $\boldsymbol{R_{BI}}$ is the rotation matrix from the body-fixed frame to the geocentric inertial frame.The dynamical state of the system is described by both the translational motion of the center of mass and the rotational motion about it. The state variables are [$\boldsymbol{r_c}$, $\boldsymbol{v_c}$, $\boldsymbol{\omega}$, $\boldsymbol{q}$ ], where $\boldsymbol{q}$ represents the attitude quaternion and $\boldsymbol{\omega}$ is the angular velocity projected onto the asteroid's body-fixed frame. The equations of motion are as follows:

$$\dot{\boldsymbol{r}}_c = \boldsymbol{v}_c \tag{3}$$

$$\dot{\boldsymbol{v}}_c = \frac{\boldsymbol{F_{ext}}}{M} \tag{4}$$

$$\dot{\boldsymbol{q}} = \frac{\boldsymbol{1}}{\boldsymbol{2}}\boldsymbol{\Omega(\omega)q} \tag{5}$$

$$\mathbf{I}\dot{\boldsymbol{\omega}} + \boldsymbol{\omega} \times (\mathbf{I}\boldsymbol{\omega}) = \boldsymbol{\tau}_{ext} \tag{6}$$

Here, **I** is the inertia tensor matrix in the body-fixed frame, and $\boldsymbol{\Omega(\omega)}$ is the quaternion multiplication matrix associated with the angular velocity. We employ a fourth-order Runge-Kutta method to numerically integrate this dynamical system.

### 2.4 Inversion Methodology

Our objective is to utilize simulated observational data (orbital and spin states) generated by the forward dynamics model to invert the low-order internal mass structure parameters of the asteroid, namely the six independent parameters of the inertia tensor matrix($I_{xx}$, $I_{yy}$, $I_{zz}$, $I_{xy}$, $I_{xz}$, $I_{yz}$)and the position vector of the center of mass in the body-fixed frame $(\rho_x, \rho_y, \rho_z)^T$ . The inversion is performed by decoupling the orbital and rotational motions.

#### 2.4.1 Inversion for the Moment of Inertia Based on the Dynamic Equation

We start from the rigid-body attitude dynamics, specifically the Euler equations:

$$\mathbf{I}\dot{\boldsymbol{\omega}} + \boldsymbol{\omega} \times (\mathbf{I}\boldsymbol{\omega}) = \boldsymbol{\tau}_{ext} \tag{7}$$

Given that precise gravitational force and torque information cannot be directly obtained via spherical harmonic expansion or the polyhedron method when the internal mass structure is unknown, we compute an approximation of the external torque based on the relationship between the dynamical state parameters and the gravitational torque. For a rigid body, the principal moment of the gravitational force about its center of mass $\boldsymbol{M_o}$ can be expressed as:

$$\boldsymbol{M}_o = \int \boldsymbol{\rho} \times \boldsymbol{dF} = -GM\int \frac{\boldsymbol{\rho} \times \boldsymbol{r}}{\boldsymbol{r}^3} dm \tag{8}$$

where $\boldsymbol{\rho}$ denotes the radius vector from the center of mass to a volume element in the body-fixed frame, $\boldsymbol{r}$ is the position vector of that element in the geocentric inertial frame, and $M$ is the total mass. By performing a Taylor expansion on the $\boldsymbol{1/r^3}$ term and substituting it into the integral for calculation, approximate expressions for the projections of the gravitational torque onto the axes of the body-fixed frame can be derived:

$$M_x = \frac{3GM}{R^3}(C-B)a_{32}a_{33} \tag{9}$$

$$M_y = \frac{3GM}{R^3}(A-C)a_{31}a_{33} \tag{10}$$

$$M_z = \frac{3GM}{R^3}(B-A)a_{32}a_{31} \tag{11}$$

where A, B, C are the moments of inertia about the asteroid's three principal axes (A > B > C), and $a_{31} = \sin\varphi\sin\theta,\ a_{32} = \cos\varphi\sin\theta, a_{33} = \cos\theta$ are the corresponding elements of the transformation matrix between the body-fixed and inertial frames. Converting the obtained gravitational torque components into the translational frame yields the approximately calculated torque $\boldsymbol{\tau}_{ext}$ , which is then substituted into the Euler dynamical Eq. (7). We numerically compute the angular acceleration $\dot{\boldsymbol{\omega}}(t)$ from the observed angular velocity data $\boldsymbol{\omega}(t)$ using a fifth-order central difference method. Substituting the known state parameters into the dynamics equation and rearranging it into a residual form concerning the inertia matrix parameters yields:

$$J(\theta_I) = \left\| \dot{\boldsymbol{\omega}} + \boldsymbol{\omega} \times (\boldsymbol{I\omega}) - \boldsymbol{\tau}_{ext} \right\| \tag{12}$$

By minimizing the residual $J(\theta_I)$ over the observation period, the moment of inertia can be solved. It should be noted that both the Euler equations and the torque approximation formula remain invariant under linear scaling transformations of the inertia matrix. The dynamical response is sensitive only to the relative ratios of the moments of inertia. Therefore, the inversion yields a normalized inertia matrix.

**2.4.2 Orbit-Fitting for Center-of-Mass Position**

After solving for the moment of inertia matrix using the attitude dynamics equations, we proceed to analyze the influence of the internal mass distribution on the offset of the center-of-mass position by studying the orbital motion of Apophis.

For a point mass, its motion within Earth's gravitational field strictly follows a Keplerian orbit. Given that Apophis's dimensions are significantly smaller than both Earth's radius and the orbital altitude, and assuming its mass is concentrated at the center of mass, we can neglect the dynamic coupling between the translational motion of the center of mass and the attitude motion induced by Earth's gravity. Thus, the motion of the center of mass can be approximated as a Keplerian orbit. We achieve the inversion of the center of mass's relative position in the body-fixed frame, $\boldsymbol{\rho_c}=(\rho_x,\rho_y,\rho_z)^T$ ,by fitting an ideal Keplerian orbit. The observed position and velocity of a feature point, $\boldsymbol{r_o}$ and $\boldsymbol{v_o}$, are converted into estimated states of the center of mass using the current attitude and the unknown $\boldsymbol{\rho_c}$:

$$\boldsymbol{r}_c^{obs} = \boldsymbol{r}_o + \boldsymbol{R}_{BI} \cdot \boldsymbol{\rho}_c \tag{13}$$

$$\boldsymbol{v}_c^{obs} = \boldsymbol{v}_o + \boldsymbol{R}_{BI}(\boldsymbol{\omega} \times \boldsymbol{\rho}_c) \tag{14}$$

In the two-body problem, the motion of the center of mass obeys Newton's law of universal gravitation, and its state equation is:

$$\begin{cases} \dfrac{d\boldsymbol{r}_c}{dt} = \boldsymbol{v}_c \\ \dfrac{d\boldsymbol{v}}{dt} = -\dfrac{GM}{\|\boldsymbol{r}\|}\boldsymbol{r}_c^3 \end{cases} \tag{15}$$

Starting with the initial center-of-mass state from Eq. (13), we numerically integrate the Keplerian orbit described by Eq. (15) to obtain the predicted center-of-mass trajectory $\boldsymbol{r}_c^{pre}(t)$.The objective function is defined as the discrepancy between the estimated and predicted Keplerian trajectories over the observation period:

$$J(\boldsymbol{\rho}) = \sum_{k=1}^{N} \left\| \boldsymbol{r}_c^{pre}(t_k) - \boldsymbol{r}_c^{obs}(t_k) \right\|_2^2 \tag{16}$$

The optimal $\boldsymbol{\rho}_c$ is solved by minimizing $J(\rho)$ .If $\boldsymbol{\rho}_c$ points to the correct center-of-mass position, then $\boldsymbol{r}_c^{obs}$ conforms to an ideal Keplerian orbit. Otherwise, it will exhibit "wobbles" induced by the center-of-mass offset, causing deviations from $\boldsymbol{r}_c^{pre}$.

**2.5 Particle Swarm Optimization**

The inversion problems described above are both highly nonlinear and degenerate. PSO enables an effective search for the global optimum across the entire solution space through information sharing and cooperation among particles, thereby avoiding entrapment in local minima. We first invert for the inertia matrix **I**, then fix **I** to invert for $\boldsymbol{\rho}_c$. Fig. 1 illustrates the basic PSO-based inversion flowchart.

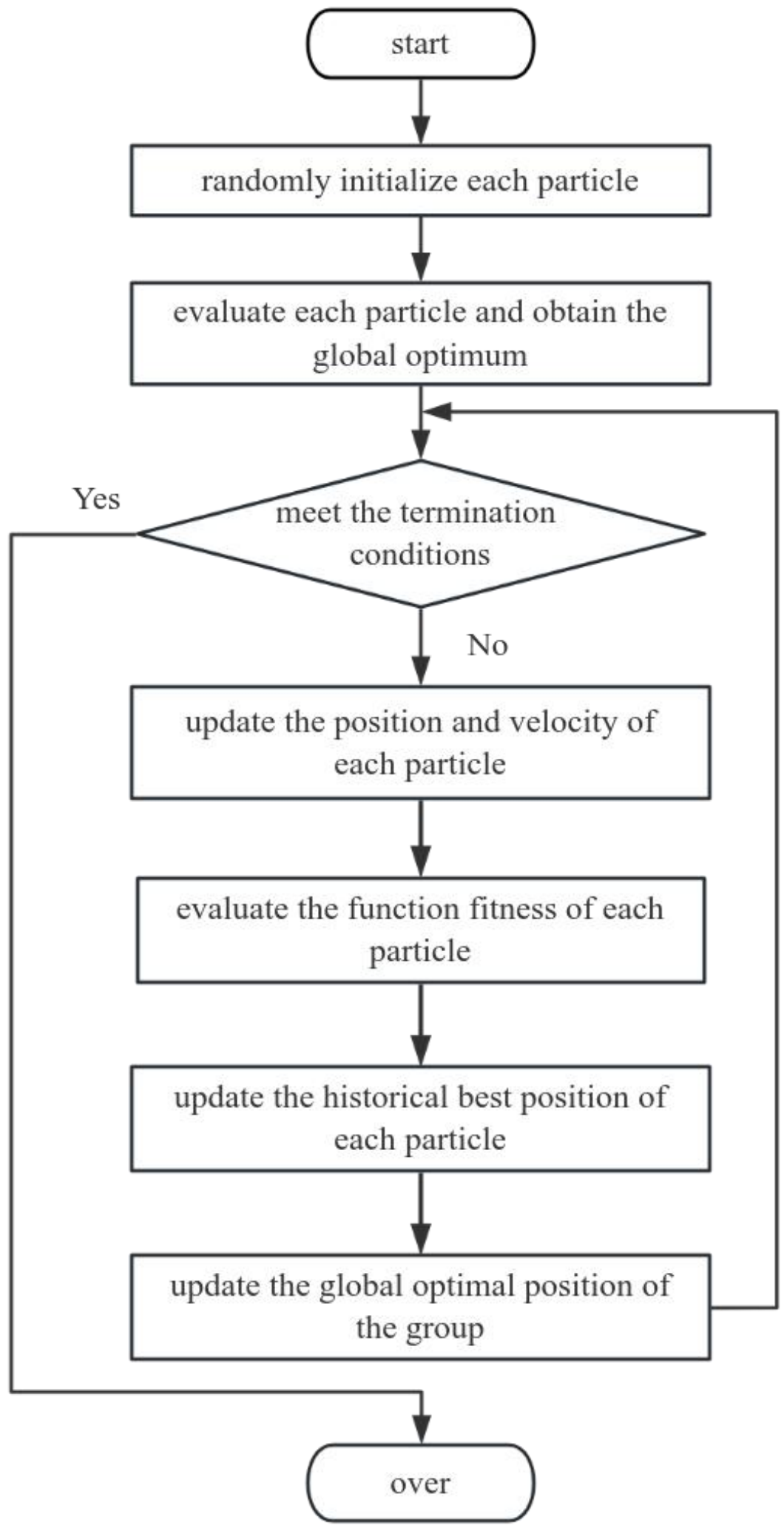

**Figure 1.** Schematic diagram of the PSO algorithm workflow.

Assume a swarm contains particles moving within the solution space. The position of the i-th particle at iteration is denoted as $\boldsymbol{x_i}=(x_1,x_2,\ldots,x_D)$, and its velocity as $\boldsymbol{v_i}=(v_1,v_2,\ldots,v_D)$. Each particle's position is substituted into a fitness function to compute its corresponding fitness value. The velocity update formula for each iteration is:

$$\boldsymbol{v}_i^{t+1} = w\cdot\boldsymbol{v}_i^t + c_1 r_1^t(\boldsymbol{pBest}_i^t - \boldsymbol{x}_i^t) + c_2 r_2^t(\boldsymbol{gBest}_i^t - \boldsymbol{x}_i^t) \quad (17)$$

We set the number of particles to 50, initialized the inertia weight $w$ to 0.8, and linearly decreased it to 0.4 during the iteration process. This grants the algorithm stronger global exploration capability in the early stages and gradually enhances local search ability later on. Both the individual and social learning factors, $c_1$ and $c_2$, are set to 1.5. The maximum number of iterations is set to 2000. The particle's position is updated based on its velocity:

$$\boldsymbol{x}_i^{t+1} = \boldsymbol{x}_i^t + \boldsymbol{v}_i^{t+1} \quad (18)$$

The algorithm continues iterating until convergence is achieved or the maximum number of iterations is reached.

**3 Results**

### 3.1 Model Analysis Test

We first constructed a regular triaxial ellipsoid model to verify the core logic and computational accuracy of the inversion algorithm. The geometric dimensions and mass of the model were set with reference to estimated values for Apophis (Otegi et al. 2020). The interior of the ellipsoid was uniformly filled using a mass-point cloud to generate two distinct internal structures: one with a uniform mass distribution and another with a density that decreases radially from the center outward. Using the numerical model described in Chapter 2, we performed a 100,000-second numerical integration of the orbital and rotational motion of the ellipsoid during a single close Earth flyby, with a fixed time step of 5 seconds, to generate an ideal "observational" data sequence. Table 2 lists the basic physical parameters of the geometric model and the initial dynamical states set for the simulation.

**Table 2.** Ellipsoid Parameters and Initial Motion State:The initial orbital state references JPL ephemeris predictions for the Apophis 2029 flyby.

| Parameter | Value |
|---|---|
| Triaxial Dimensions (m) | (120, 250, 170) |
| Total Mass (kg) | $4.6\times10^{10}$ |
| Initial Distance $r$ (km) | $5.2504\times10^{8}$ |
| Initial Relative Velocity $v$ (m/s) | $5.0290\times10^{3}$ |
| Angular Velocity $\omega$ (rad/s) | $5.82\times10^{-5}$ |

The simulated orbital and rotational data served as input for the inversion algorithm to solve for the moment of inertia and center-of-mass information, respectively. In each inversion process, 50 particles were used in parallel for optimization. The initial particles for the moment of inertia parameters ($I_{xx}$, $I_{yy}$, $I_{zz}$, $I_{xy}$, $I_{xz}$, $I_{yz}$) were randomly generated within a 68% confidence interval centered on the true values. The initial guess for $(\rho_x, \rho_y, \rho_z)^T$ was randomly initialized within the ellipsoid's volume. The fitness of each particle was calculated according to Eqs. (12) and (16), and the corresponding inertia tensor and center-of-mass position were obtained through iterative optimization.

We conducted inversion tests using orbital and rotational data from different time spans to compare the resulting parameter accuracy. Experiments indicated that using approximately 40,000 seconds of data, starting from the initial state (Table 2), as input yielded inversion results with relatively high accuracy. Fig.2(a) to 2(e) show the convergence curves of the normalized inertia matrix parameters versus iteration number for the uniform density model. Fig.2(f) presents the corresponding convergence curve of the fitness function. The relative error of all inverted parameters was less than $10^{-7}$. Fig. 3 displays the convergence curves of the fitness function and the three coordinate components during the center-of-mass iteration process. Although the center-of-mass position of the symmetric structure is close to the geometric center, the inversion algorithm successfully identified the minute offset with a precision on the order of $10^{-8}$, demonstrating the effectiveness of the method.

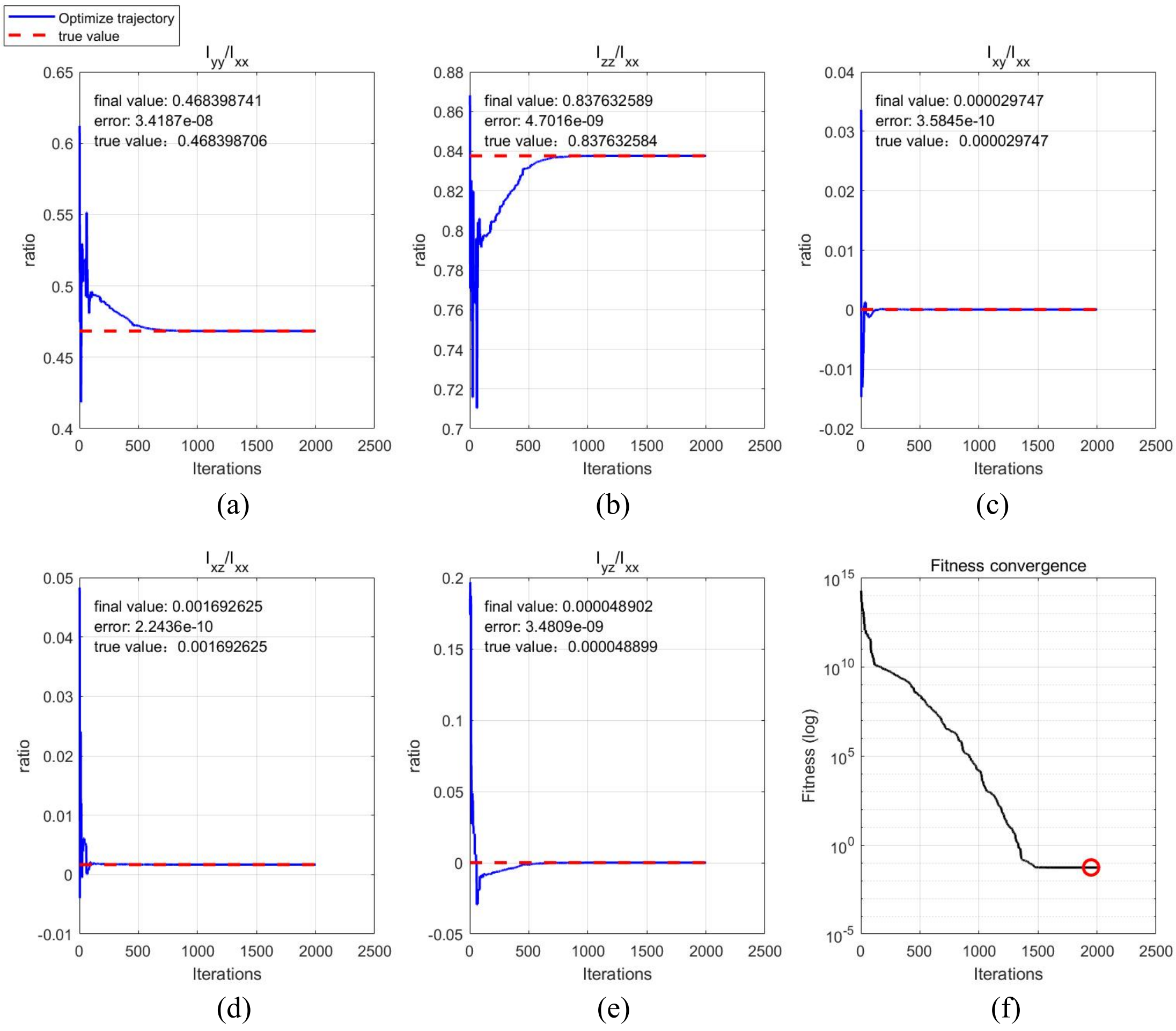

(a) (b) (c)

(d) (e) (f)

**Figure 2.** Inversion iteration results for the inertia matrix parameters of the uniform-density ellipsoid model. The red lines indicate the normalized true proportional parameters, and the blue lines show the convergence curves of the parameters.

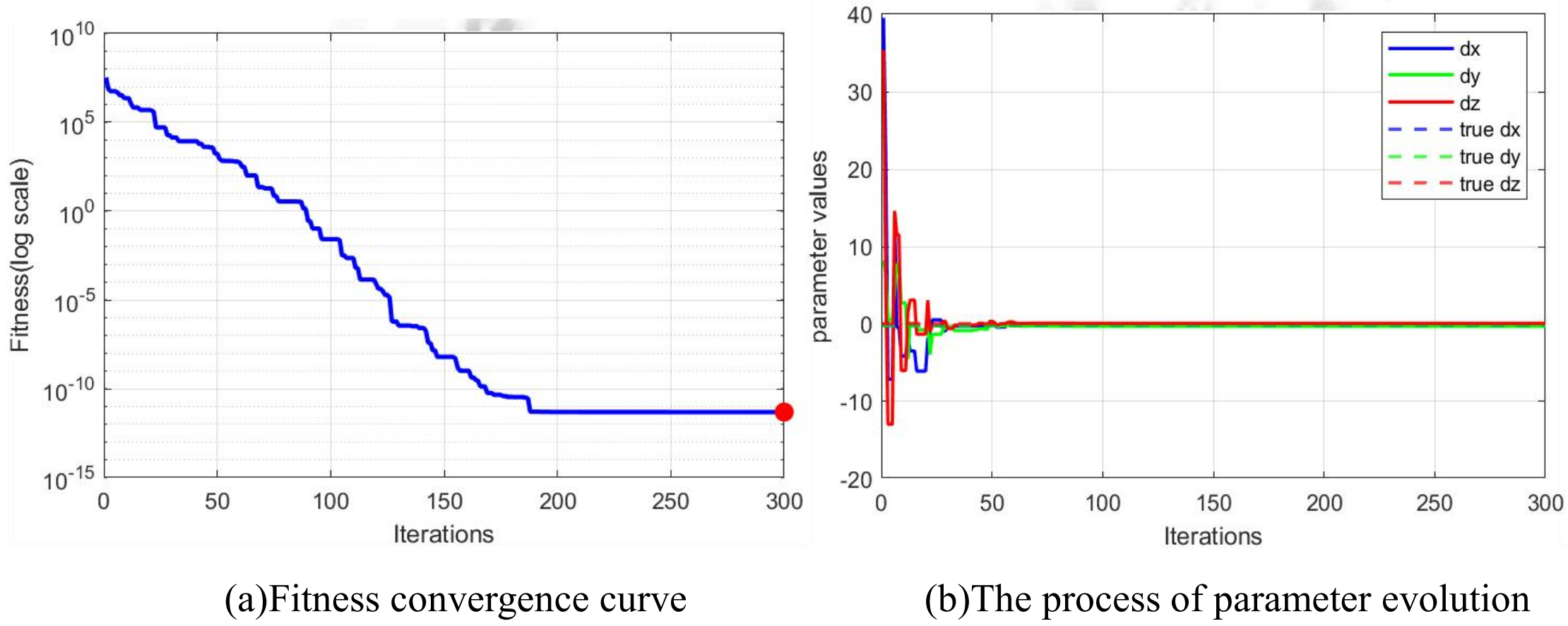

(a)Fitness convergence curve (b)The process of parameter evolution

**Figure 3.** Inversion iteration results for the center-of-mass position of the uniform-density ellipsoid model: (a) shows the convergence process of the fitness curve, which stabilizes after approximately 180 iterations; (b) shows the iterative

convergence process of the three coordinate components of the center of mass.

For the model with a radially decreasing density distribution, the ratios of the principal moments of inertia, $I_{yy}/I_{xx}$ and $I_{zz}/I_{xx}$, show noticeable differences compared to the uniform model. Figs. 4 and 5 present the convergence curves for the moment of inertia and the center of mass of this model, respectively. The inversion accuracy is similar to that of the uniform model, preliminarily indicating that the algorithm can effectively distinguish and accurately solve for the corresponding low-order mass parameters under different internal density configurations. The PSO algorithm converges within a similar number of iterations.

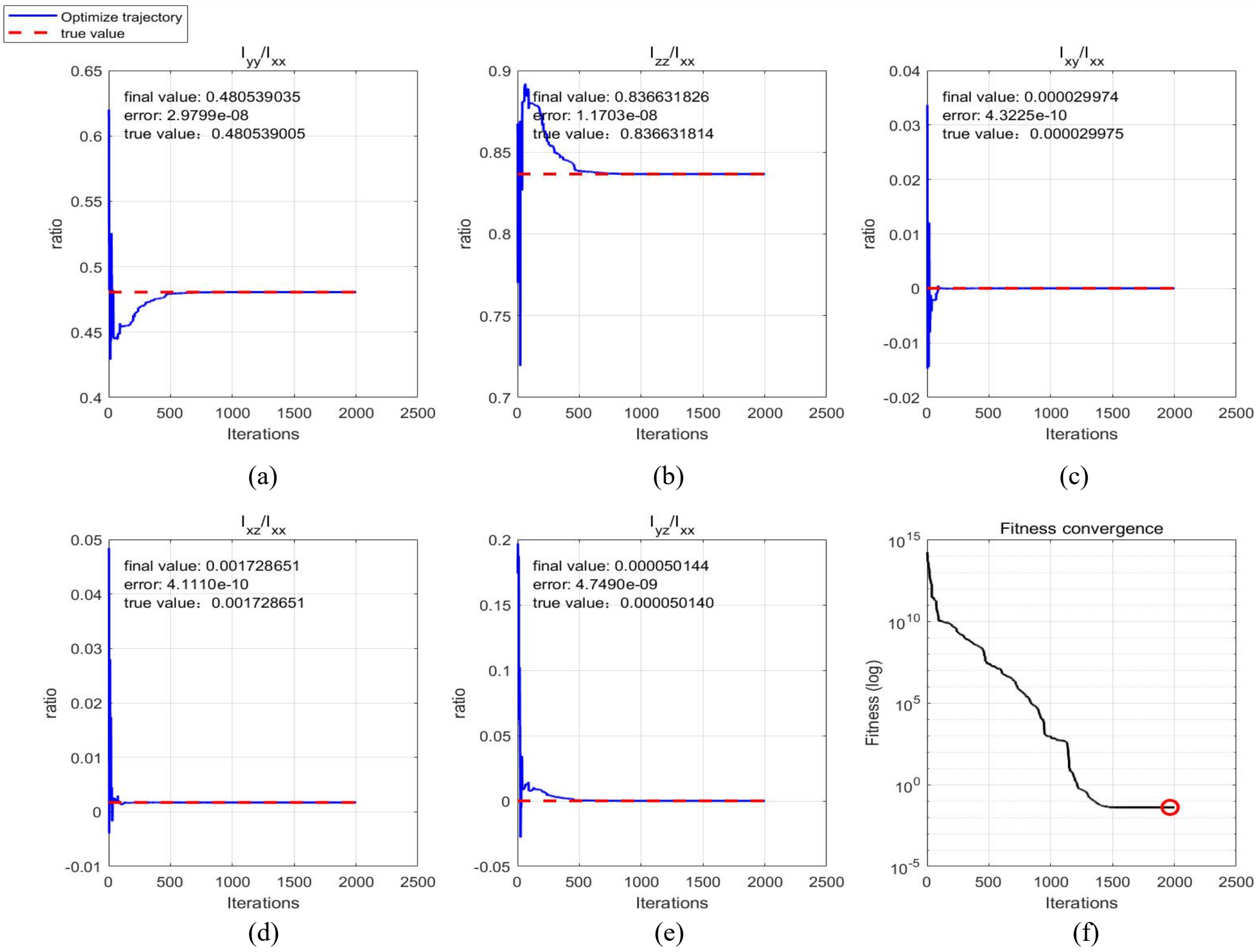

**Figure 4.** Iteration results for the inertia matrix parameters of the ellipsoid model with radially varying density. The red lines represent the normalized true proportional parameters, and the blue lines show the parameter convergence curves. The descent process of the fitness curve is similar to that of the uniform model. As the major axis of the ellipsoid aligns with the y-axis, the difference in $I_{yy}/I_{xx}$ is most pronounced compared to the uniform model.

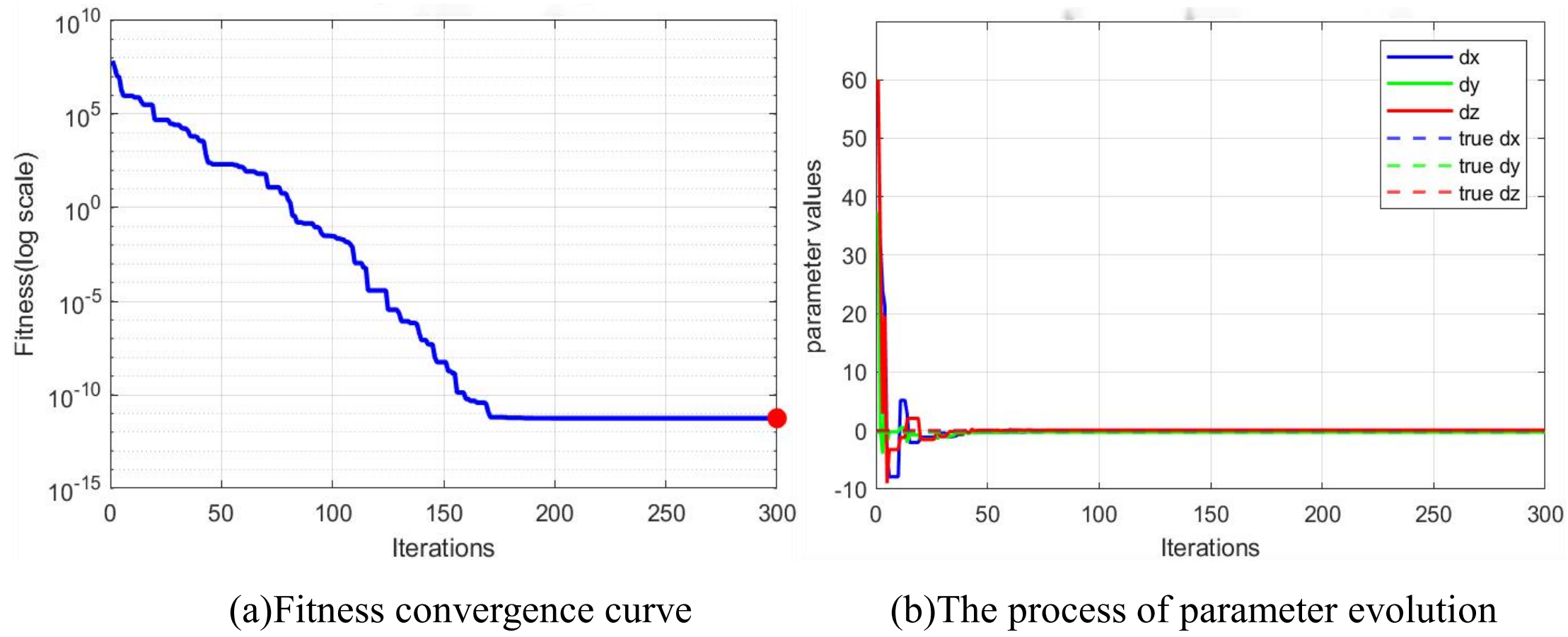

(a)Fitness convergence curve (b)The process of parameter evolution

**Figure 5.** Iteration results for the center of mass of the ellipsoid model with radially varying density: (a) shows the convergence process of the fitness curve, which stabilizes after approximately 180 iterations; (b) shows the iterative process of the three coordinate components of the center of mass.

**3.2 Simulation Verification for Apophis**

To systematically evaluate and demonstrate the performance of the proposed internal mass structure inversion method in a practical application scenario, we constructed an irregular shape model based on the predicted 2029 flyby trajectory of Apophis and its existing 3D point cloud data. Within this shape model, we defined three distinct internal mass distribution models, each representing a different scientific hypothesis, to test the algorithm:

*Homogeneous Model:* All mass points are assigned equal mass, serving as a baseline reference. *"Bilobed" Model:* Emulating the contact binary formation hypothesis, the average density of mass points in the right hemisphere (along the positive x-axis direction) is set 30% higher than that in the left hemisphere. *"Rubble-pile" Model:* Multiple low-density spherical void regions, each approximately 10% of the asteroid's size, are embedded within the interior to simulate porosity. The density of mass points in the central region (within 40% of the asteroid's size) is set 30% higher than in the outer region.

Fig. 6 shows the 3D shape mesh of Apophis used in this study, with its geometric center marked by a red dot.

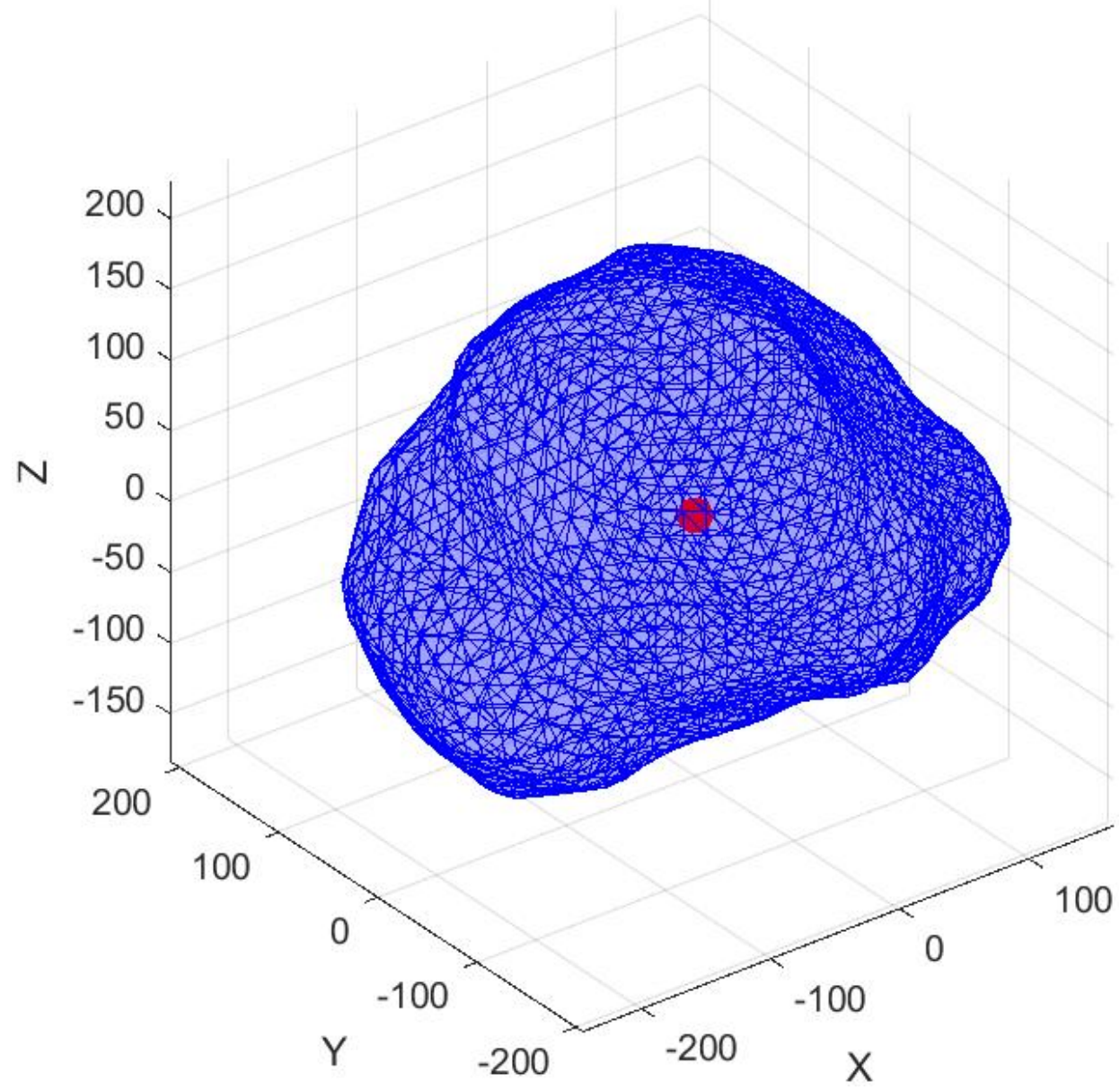

**Figure 6.** Schematic of the Apophis shape. The red solid circle indicates the geometric center.

Fig. 7 presents examples of the three different internal mass distribution models generated based on the Apophis shape. The total mass for all models is $4.60 \times 10^{10}$kg.

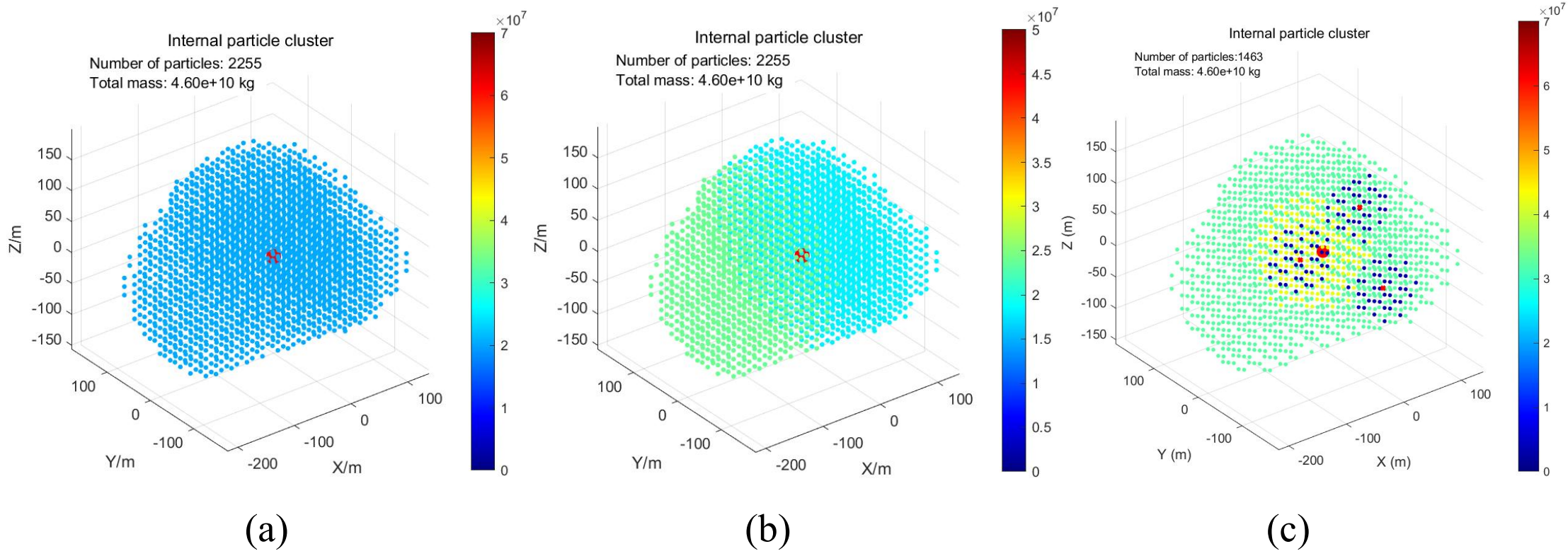

(a) (b) (c)

**Figure 7.** Examples of different mass distribution models for Apophis: **(a) Homogeneous model:** Total mass is uniformly distributed among all mass points. **(b) Bilobed model:** The density of all mass points in the right hemisphere is 30% higher than in the left hemisphere along the x-axis direction. **(c) "Rubble-pile" model:** The density of mass points in the central region is 30% higher than in the outer region, with three very low-density spherical zones placed near the center to simulate internal voids. The color bar represents the mass of individual mass points.

Using the initial orbital state from Table 2, numerical orbit and spin integration was performed for each model to generate simulated observation datasets. Taking the most complex "rubble-pile" model as an example, its simulated observation data spanning 40,000 seconds were fed into the inversion algorithm. Fig. 8 illustrates the convergence process of the key ratio parameters of the inertia matrix for this model. Each subplot shows the optimization trajectory of one of the five normalized moment of inertia ratios, along with the change in the fitness value corresponding to the global best particle across iterations. In the initial 500 iterations, the algorithm is in a rapid global exploration phase, with significant fluctuations in parameter values and fitness.

Subsequently, it enters a local refinement phase, where the parameter trajectories converge stably, and the fitness value monotonically decreases to the order of $10^{-12}$ and stabilizes.

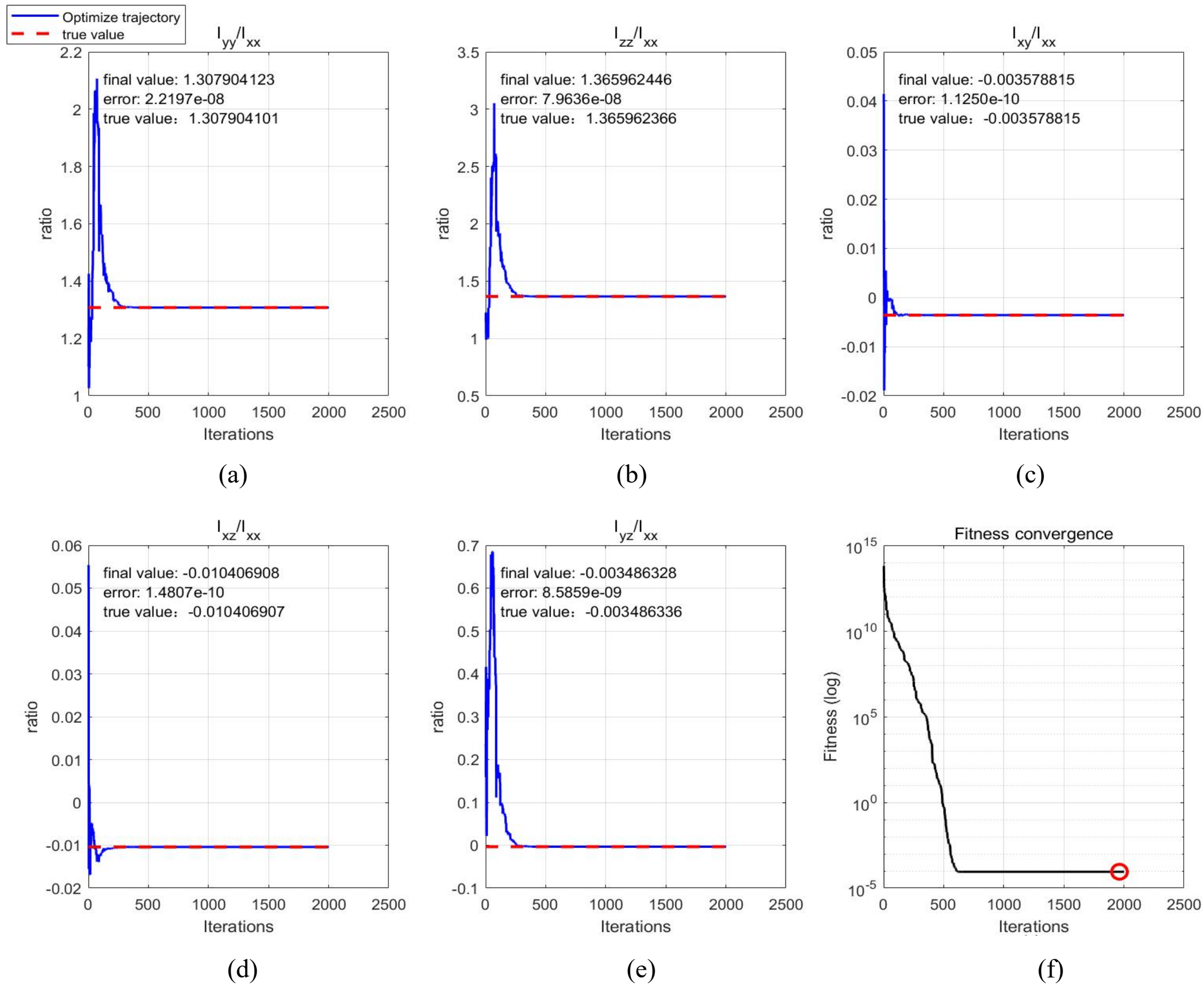

(a) (b) (c)

(d) (e) (f)

**Figure 8.** Convergence process of the inertia tensor parameter inversion for the rubble-pile model: Subfigures (a)-(e) show the optimization trajectories of the normalized moment of inertia ratio parameters $I_{yy}/I_{xx}$、$I_{zz}/I_{xx}$、$I_{xy}/I_{xx}$、$I_{xz}/I_{xx}$、$I_{yz}/I_{xx}$, with red dashed lines indicating their true values. (f) The descent curve of the fitness function versus the number of iterations, demonstrating stable and rapid convergence of the algorithm.

The inversion process for the center-of-mass coordinates is shown in Fig. 9. The particles corresponding to the three coordinate components converge quickly towards their true values in the early iterations and gradually stabilize.

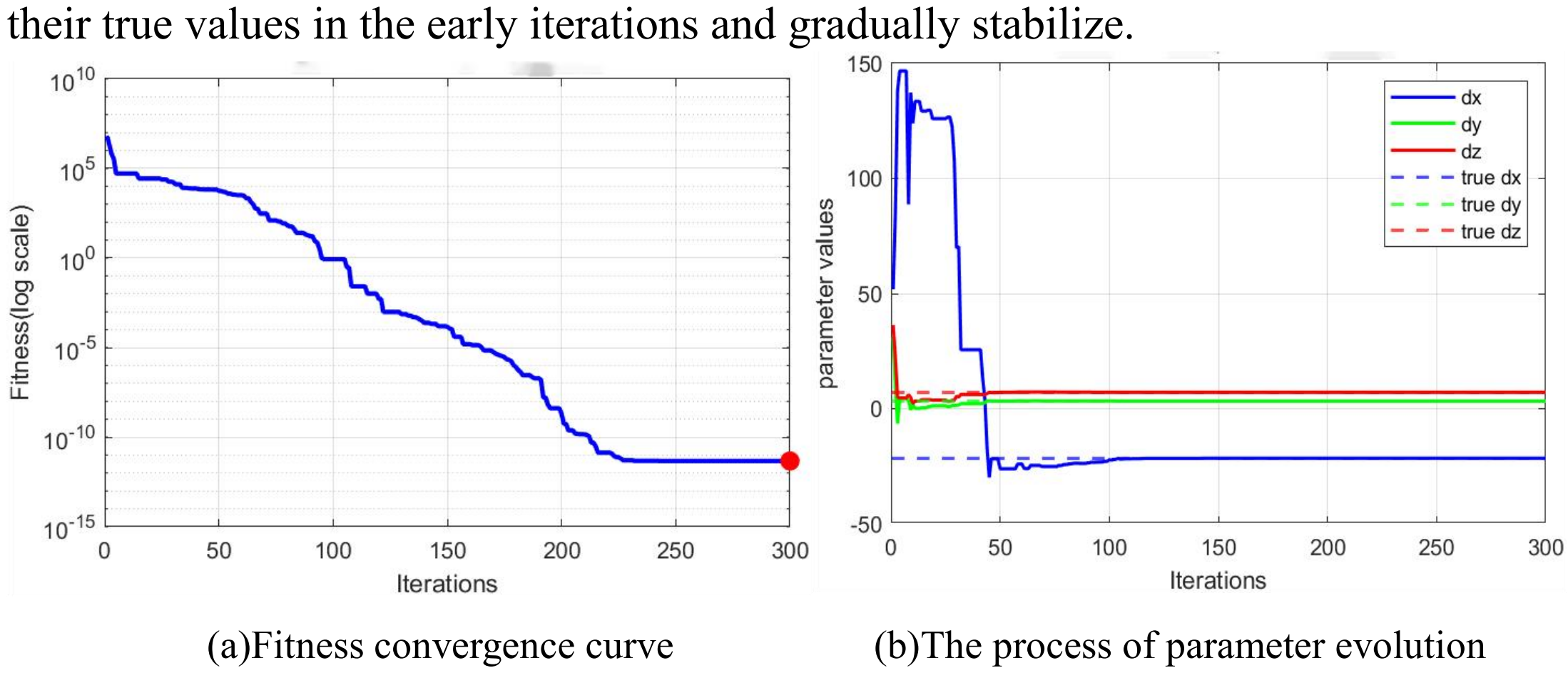

(a)Fitness convergence curve (b)The process of parameter evolution

**Figure 9.** Evolution of the center-of-mass coordinates for the rubble-pile model. The dashed lines in (b) represent the true values of the three coordinate components, respectively.

The inversion results for the three Apophis mass distribution models under ideal conditions are summarized in Table 3. The table lists the ratio parameters of the normalized inertia matrix, as well as the true values, inverted values, and their errors for the center-of-mass coordinate components in the body-fixed frame. For each model, the relative errors of all inertia matrix ratio parameters are less than 0.001%, and the absolute errors of the center-of-mass coordinate components are less than 5 × $10^{-6}$ meters.The differences in the inertia ratios across the three models are not merely numerical, they reflect the underlying dynamical characteristics.

**Table 3.** Inversion results for the three internal mass distribution models of Apophis.

| Parameter | Homogeneous Model | | Bilobed Model | | Rubble-pile Model | |
|---|---|---|---|---|---|---|
| **Moment of Inertia Ratios** | **True Value** | **Inverted Value (Rel. Error)** | **True Value** | **Inverted Value (Rel. Error)** | **True Value** | **Inverted Value (Rel. Error)** |
| $I_{yy}/I_{xx}$ | 1.2984278 | 1.2984278 ($<1\times10^{-8}$) | 1.3443052 | 1.3443052 ($<1\times10^{-8}$) | 1.307904 | 1.307904 ($<1\times10^{-8}$) |
| $I_{zz}/I_{xx}$ | 1.3632469 | 1.3632469 ($<1\times10^{-8}$) | 1.3959833 | 1.3959832 ($<1\times10^{-8}$) | 1.365962 | 1.365962 ($<1\times10^{-8}$) |
| $I_{xy}/I_{xx}$ | 0.0005717 | 0.0005717 ($<1\times10^{-8}$) | 0.0004495 | 0.0004495 ($<1\times10^{-8}$) | -0.003578 | -0.003578 ($<1\times10^{-8}$) |
| $I_{xz}/I_{xx}$ | 0.0053567 | 0.0053567 ($<1\times10^{-8}$) | 0.0063184 | 0.0063184 ($<1\times10^{-8}$) | -0.010406 | -0.010406 ($<1\times10^{-8}$) |
| $I_{yz}/I_{xx}$ | 0.0035014 | 0.0035014 ($\approx7\times10^{-7}$) | 0.0093341 | 0.0093341 ($<1\times10^{-8}$) | -0.003486 | -0.003486 ($\approx2.5\times10^{-6}$) |
| **Center of Mass(m)** | **True Value** | **Inverted Value (Abs. Error)** | **True Value** | **Inverted Value (Abs. Error)** | **True Value** | **Inverted Value (Abs. Error)** |
| $\rho_x$ | -19.899574 | -19.899574 ($4.322\times10^{-6}$) | -30.9161850 | -30.9161840 ($1.043\times10^{-6}$) | -21.5233557 | -21.52335261 ($1.611\times10^{-7}$) |
| $\rho_y$ | -0.856667 | -0.8566670 ($-2.428\times10^{-8}$) | -0.7891894 | -0.7891895 ($-7.282\times10^{-9}$) | 3.3382022 | 3.33820217 ($-1.091\times10^{-9}$) |
| $\rho_z$ | 8.565025 | 8.56502508 ($-3.557\times10^{-8}$) | 8.3571453 | 8.3571453 ($-8.888\times10^{-9}$) | 7.1189639 | 7.11896394 ($-1.612\times10^{-9}$) |

We first consider the homogeneous model serves as a baseline. Its inertia ratios are a direct consequence of the shape's geometry.The values $I_{yy}/I_{xx}$=1.2984 and $I_{zz}/I_{xx}$=1.3632 reflect that the longest dimension is along the x-axis (minimum inertia).The small but non-zero off-diagonal terms ($I_{xy}/I_{xx}$, $I_{xz}/I_{xx}$, $I_{yz}/I_{xx}$) arise from the fact that the principal axes of the shape model do not perfectly align with the body-fixed coordinate axes used in our discretization.

The bilobed model concentrates 30% higher density in the right hemisphere along the x-axis. Compared to the homogeneous case, we observe a marked increase in $I_{yy}/I_{xx}$ (+3.5%) and $I_{zz}/I_{xx}$ (+2.4%). This is expected: shifting mass toward one lobe along the x-axis increases the moments of inertia about the perpendicular y- and z-axes. More strikingly, the off-diagonal term $I_{yz}/I_{xx}$ increases from 0.0035 (relative to homogeneous) to 0.0093, indicating that the mass asymmetry rotates the principal axes away from the geometric axes.

The rubble-pile model features a central region that is 30% denser than the outer shell, together with three low-density voids. Its inertia ratios lie close to the homogeneous case, but the off-diagonals become negative. These negative values capture the orientation of the internal inhomogeneities relative to the body-fixed frame, and their magnitudes quantify the degree of “dynamical asymmetry” introduced by the internal structure.This demonstrates that the method can effectively detect and quantify the asymmetry of the inferred inertia tensor, which reflects the deviation of the internal mass distribution from the principal axes.

### 3.3 Observation Uncertainty Impact and Sensitivity Analysis

In practical observations, due to limitations in the accuracy and resolution of measurement instruments such as radar and sensors, observational data inevitably contain noise, which introduces deviations into the final optimization results. To assess the feasibility of the proposed method in actual missions, we introduce Gaussian noise into the simulated data based on the survey of existing observational technologies and expected accuracy conditions outlined in Section 2.2. Specifically, the relative noise for position and velocity is set to $1\times10^{-7}$, for attitude observations to $1\times10^{-4}$, and for angular velocity to $1\times10^{-7}$. These values are conservative estimates based on state-of-the-art radar and optical tracking capabilities (Brozović et al. 2018; Naidu et al. 2016). We perform 100 independent inversion runs for each noise configuration and analyze the statistical distributions of the inverted parameters. Figure 10 shows the histograms and fitted normal distributions of the five normalized inertia ratios obtained from 100 noisy-data inversions for the rubble-pile model. Under the baseline noise levels, the uncertainty in the inversion of each moment of inertia parameter can be controlled within approximately 0.1%. The off-diagonal terms exhibit larger relative errors than the diagonal ratios.This is expected because off-diagonal terms are typically smaller in magnitude and thus more susceptible to noise contamination. The inversion of inertia parameters relies on the Euler equation: These equations directly link the time derivative of angular velocity to the external torque $\boldsymbol{M_o}$ via the inertia tensor $\mathbf{I}$. To compute the residual in Eq.12, we must numerically differentiate the observed $\boldsymbol{\omega}(t)$ to obtain $\dot{\boldsymbol{\omega}}$. Even small relative noise in $\boldsymbol{\omega}$ can be amplified by the differentiation process, leading to non-negligible errors in $\dot{\boldsymbol{\omega}}$. By contrast, errors in position and attitude affect the external torque $\boldsymbol{M_o}$ only indirectly through the Earth–asteroid relative geometry.

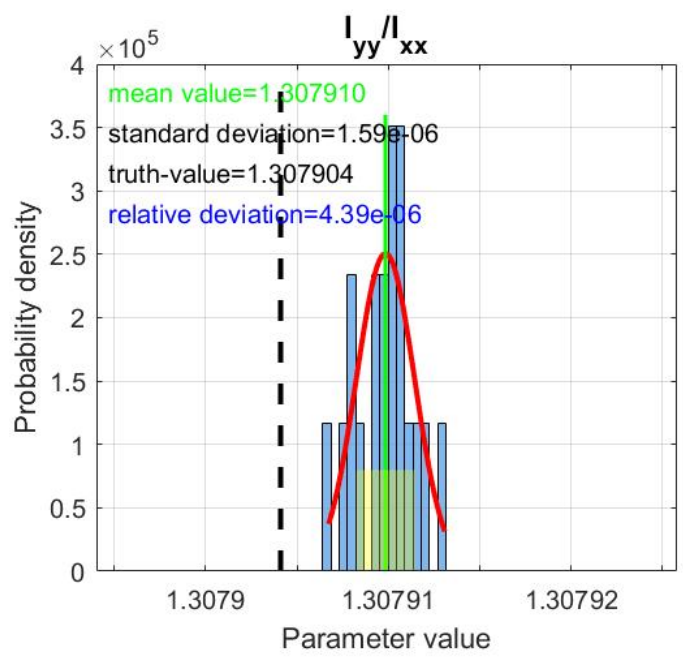

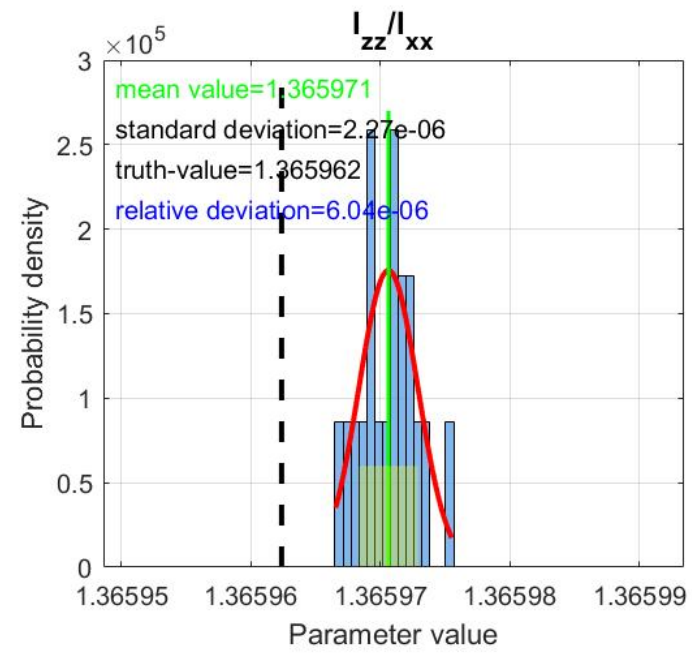

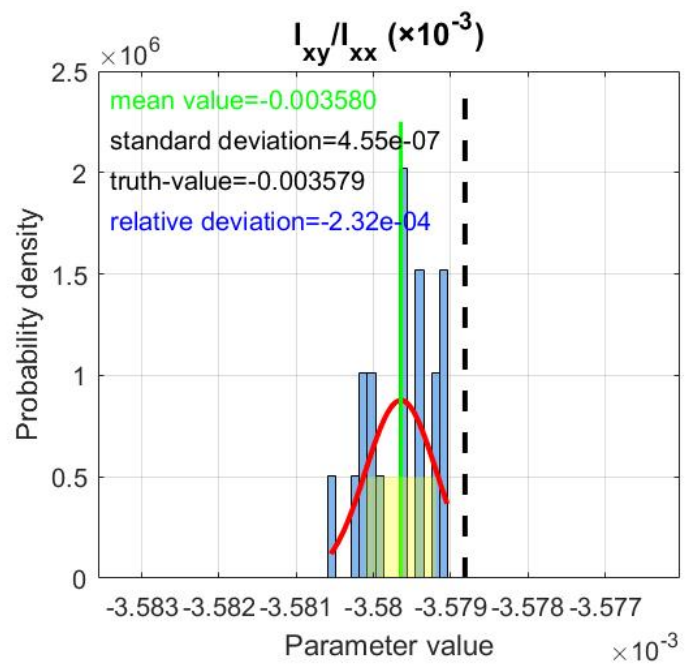

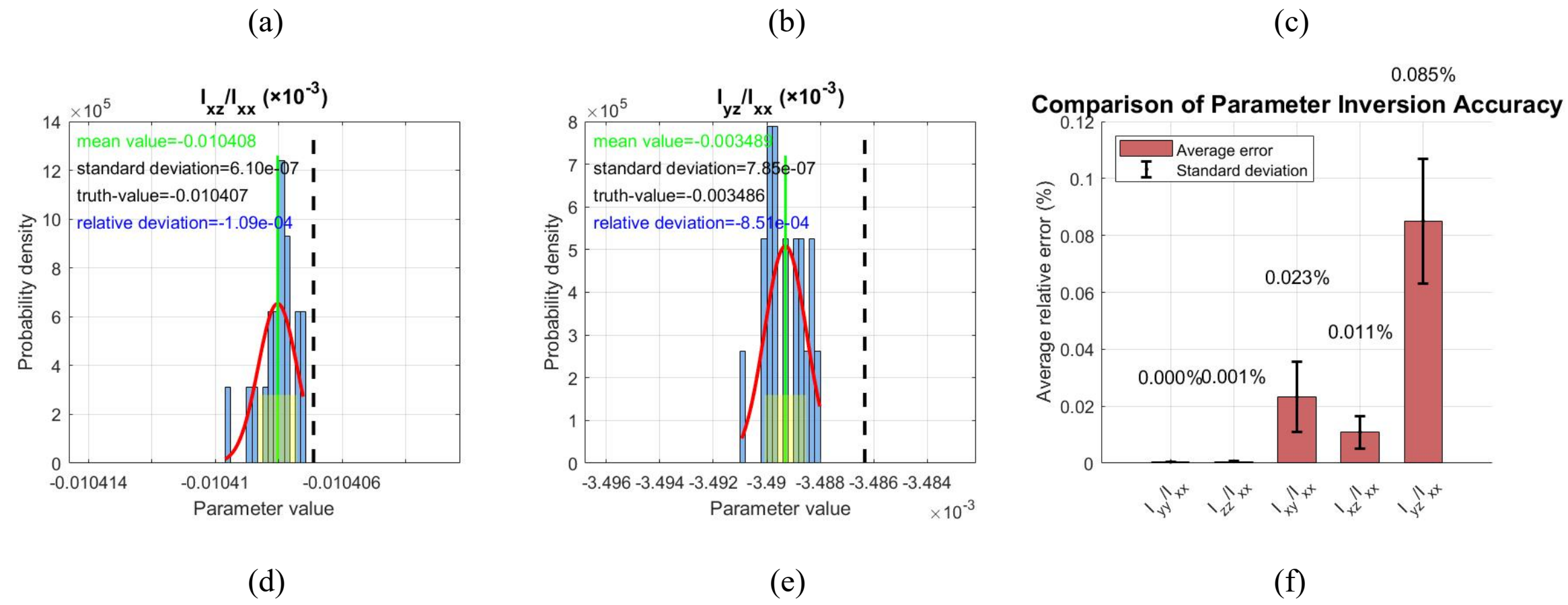

**Figure 10.** Statistical distribution of the inversion results for the moment of inertia ratio parameters of the "rubble-pile" model. Subfigures (a)-(e) display the distribution of the inversion results for the normalized moment of inertia parameters $I_{yy}/I_{xx}$、$I_{zz}/I_{xx}$、$I_{xy}/I_{xx}$、$I_{xz}/I_{xx}$、$I_{yz}/I_{xx}$, respectively. In each subfigure, the blue histogram represents the frequency distribution of the results from 100 independent inversions. The red solid line is the fitted normal distribution probability density function. The green solid line marks the true value of the parameter. The dashed line marks the mean of the inversion results. The yellow shaded area indicates the 68% confidence interval.

We conducted a control variable analysis to quantify the impact of noise from different observational data sources on the inversion results. The heatmap in Fig.11(a) shows the relative error in the inversion results for each moment of inertia parameter when noise is introduced solely in position, velocity, or angular velocity. By increasing the relative noise of different data sources by one or two orders of magnitude (Fig. 11(b) and (c)), and comparing the resulting relative errors in the inversion results, it is demonstrated that observation error in angular velocity is the primary factor contributing to uncertainty in moment of inertia inversion. Once its relative noise exceeds approximately 1×10$^{-7}$, it leads to a sharp increase in parameter inversion errors.

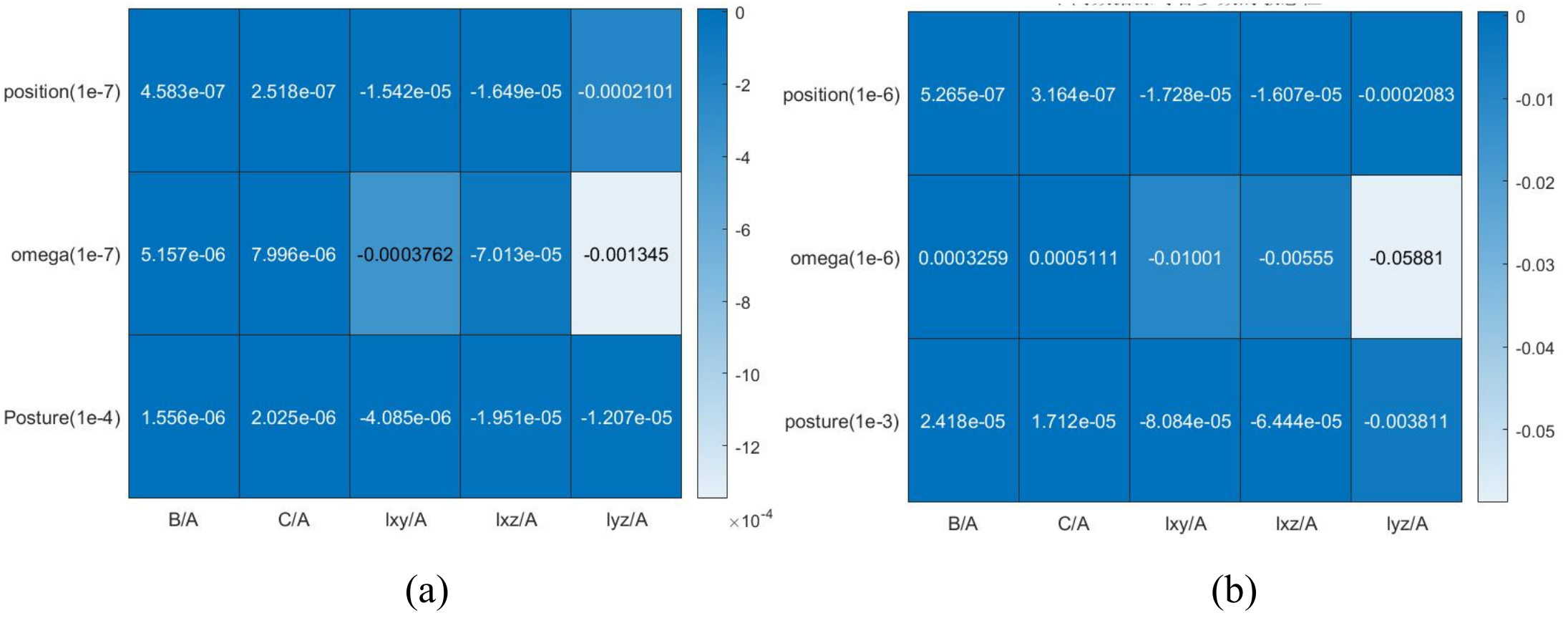

**Figure 11.** Sensitivity analysis of moment of inertia inversion under different noise levels. (a) Relative errors in the inversion results for each normalized moment of inertia ratio caused by introducing noise separately in position, attitude, or angular velocity, under baseline noise levels (Position:1×10$^{-7}$, Attitude:1×10$^{-4}$, Angular Velocity:1×10$^{-7}$). (b) Relative errors in parameter inversion results when the relative

noise from each data source is increased by one order of magnitude. The color scale represents the magnitude of the relative error.

Compared to the moment of inertia, the inversion of the center-of-mass position exhibits extreme sensitivity to noise. Even with data noise at the expected level described in Section 2.2, the center-of-mass inversion results often show significant deviations, sometimes even exceeding the physical dimensions of the asteroid, with high dispersion across multiple inversion runs. Orbital motion is a stiff system: errors in initial position or velocity produce errors that grow linearly with time (for position) or remain constant (for velocity) in the Keplerian case.The heatmap in Fig. 12 illustrates the sensitivity of the inversion error for each center-of-mass coordinate to noise in different data. The accuracy of the center-of-mass inversion is almost entirely dominated by the observation accuracy of position and velocity. To keep the inversion error below 1 meter, the observation noise for position and velocity needs to be lower than $1\times10^{-10}$ and $1\times10^{-9}$, respectively.This confirming that the proposed center of mass inversion via this method is not practically feasible with today’ s ground-based tracking alone; it would require dedicated in-situ tracking or a dramatic improvement in radar precision.

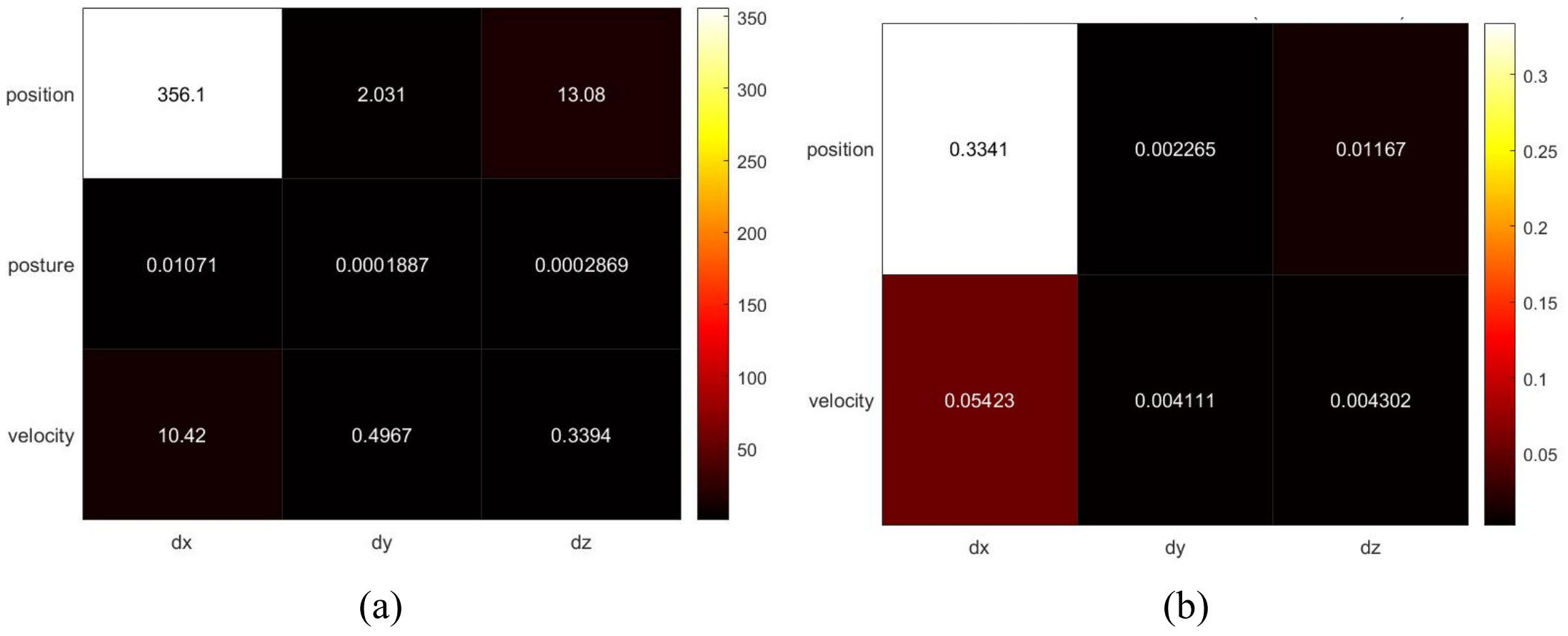

**Figure 12.** Heatmap showing the sensitivity of inversion errors for center-of-mass coordinate components to noise in different observational data. (a) Absolute errors in the inversion of each center-of-mass coordinate component when noise is introduced separately in position, attitude, or velocity, under baseline noise levels (Position/Velocity:$1\times10^{-7}$, Attitude:$1\times10^{-4}$). (b) Inversion result errors when the observation noise for position and velocity is reduced to $1\times10^{-10}$ and $1\times10^{-9}$, respectively. The color scale from dark to light indicates increasing error.

## 4. Conclusion and Discussion

This paper proposes and validates, as a proof-of-concept, an inversion method for determining an asteroid's internal mass structure parameters—specifically, the normalized inertia tensor and the COM offset within the body-fixed frame—based on its dynamical response during a close planetary flyby. Using the 2029 Earth flyby of Apophis as a representative scenario, we constructed a forward dynamics model and an inversion framework based on Particle Swarm Optimization.

Under ideal observational data conditions, for various predefined internal mass distribution models (e.g. bilobed, rubble-pile models), the inversion errors for the moment of inertia matrix parameters were all below 0.001%, and the error for the center-of-mass position was on the order of $10^{-5}$m. This confirms that, in principle, the dynamical fingerprint of a flyby contains sufficient information to uniquely recover the low-order mass distribution parameters. Compared to high-precision gravity field inversion methods that rely on long-term, close-proximity tracking observations by spacecraft, the method presented here uses only orbital and rotational data from a short time window during the flyby and requires no a priori information about the internal structure. By directly solving for the inertia matrix parameters and the center of mass, it provides constraints for assessing the non-uniformity and anisotropy of an asteroid's mass distribution and its potential formation mechanisms, offering greater timeliness and cost-effectiveness. Beyond Apophis, our method is also applicable to other near-Earth asteroids with similar flyby opportunities.

The transition from a noise-free ideal scenario to a more realistic one, however, presents significant challenges. When Gaussian noise commensurate with current or near-future observational capabilities is introduced, the inversion accuracy degrades considerably. Our sensitivity analysis clarify the data accuracy thresholds. The primary bottleneck of inertia tensor inversion is the measurement accuracy of the angular velocity. Errors in angular velocity propagate directly into the calculation of angular acceleration and the torque model, causing uncertainties in the inverted inertia ratios. To achieve reliable inversion results, the relative noise in angular velocity likely needs to be at or below the $1\times10^{-7}$ level. In comparison, the inversion is relatively robust to noise in position and attitude data.

The center of mass is an extremely sensitive parameter. Even with the baseline noise levels, the center of mass inversion results can become unstable and physically implausible, even exceeding the asteroid's dimensions. The accuracy is almost entirely governed by the precision of the orbital position and velocity data. Our analysis suggests that to constrain the center of mass error to within a few meters, the relative noise in position and velocity measurements would need to be as stringent as $1\times10^{-10}$ and $1\times10^{-9}$, respectively, which is beyond current capabilities for a fast-moving target. This represents a significant challenge for practical application.

We view it as a quantitative specification for future observation campaigns: if pinpointing the COM location is a mission objective , then in-situ tracking or a dramatic improvement in radar precision will be required. For the upcoming Apophis flyby, our results suggest that while the inertia tensor can be retrieved with confidence using existing radar and optical capabilities, determining the COM offset to better than several meters will likely remain out of reach from the ground.

Furthermore, we acknowledge that our model does not yet account for other real-world complexities such as errors in the shape model, non-continuous observation arcs, or the perturbing effects of the Sun and Moon. Future work will focus on incorporating these factors into a more robust inversion framework and exploring the use of more sophisticated optimization algorithms to move towards a more realistic scenario. This study holds significant importance not only for revealing the internal structure of Apophis and understanding its origin and evolution but also offers an economical and efficient solution for rapidly analyzing the internal properties of asteroids using their dynamic responses during flyby or impact missions. It has clear application value in the fields of asteroid science, planetary defense, and deep space exploration.

**Acknowledgements**

This work was supported by the National Natural Science Foundation of China (Grant No. 12272018, Y.Y.).